# IRA codes derived from Gruenbaum graph

Alexander Zhdanov
Voronezh, Russia
Email: a-zhdan@vmail.ru

*Abstract:*. **In this paper, we consider coding of short data frames (192 bits) by IRA codes. A new interleaver for the IRA codes based on a Gruenbaum graph is proposed. The difference of the proposed algorithm from known methods consists in the following: permutation is performed by using a match smaller interleaver which is derived from the Gruenbaum graph by finding in this graph a Hamiltonian path, enumerating the passed vertices in ascending order and passing them again in the permuted order through the edges which are not included in the Hamiltonian path. For the IRA code the obtained interleaver provides 0.7-0.8 db gain over a convolutional code decoded by Viterbi algorithm.**

## I. Introduction

Iterative decoding techniques make an alteration in forward error correction coding (FEC) where the Viterbi Algorithm dominates. But until now data frames within the interval of 24-192 bits have been coded with a convolutional code and decoded with the Viterbi algorithm. The aim of this paper is to propose a new channel algorithm for this interval. The first idea of iterative decoding was put forward in 1954 by Elias [1]. In 1963, Gallager introduced LDPC codes and the algorithm of their decoding in soft decisions [2] but these algorithms have been forgotten for thirty years due to triumph of convolutional codes under the Viterbi decoding algorithm [3]. The main feature of the Viterbi algorithm is a trellis construction that provides a property of maximum a posteriori probability to each decoded codeword. The iterative decoding technique returned in 1993 with a turbo-code [4] where the same data bits (but in different orders after permutation) were encoded by two simple convolutional encoders and decoding was performed by two decoders that iteratively provided each other with extrinsic information about received information bits. Each decoder implements the BCJR algorithm [5] (also known as MAP) that provides a property of maximum a posteriori probability to each decoded bit. The invention of turbo-codes started the era of attacks to Shannon limit, which was continued with re-invention of LDPC codes [6]. The effective method of encoding and implementation was constructed for this code [7] but the original Gallager's idea about ensemble of simple parity check decoders has not been changed. All decoding streams in both mentioned methods assumed to be statistically independent in system analysis. Actually that is not true and is only fulfilled if the frame length tends to $\infty$. So, design of finite frame permutation is a hot point in development of a coding scheme.

The duality of the above-described coding techniques is pointed out in [8]. That means the turbo-code can be represented as a parity check code. The codes combining positive properties of both types of coding are known as turbo-like codes [9]. The difference between these codes and classical LDPC codes is linear encoding complexity. The difference from the classical turbo-codes is that the trellis pattern is reduced from 16 or 8 states to 2 states and there is no need to transmit any tail bits except one. The repeat-accumulate (RA) codes appeared as a simplified ("study") case of the LDPC codes but it has been proved that they could achieve a channel capacity when the word length approaches infinity [9] and their performance is not worse than that of the turbo codes [10]. The term "accumulate" corresponds to multiplication by the polynomial $g_a = \dfrac{1}{1+D}$. I.e., parity symbols can be obtained by repeating each information bit several times, interleaving the repeated bits, and further encoding the interleaved bits by the convolutional code with the generator polynomial $g_a$. The encoding device with the polynomial $g_a$ is called an accumulator. The basic RA encoder structure is demonstrated in Fig. 1.

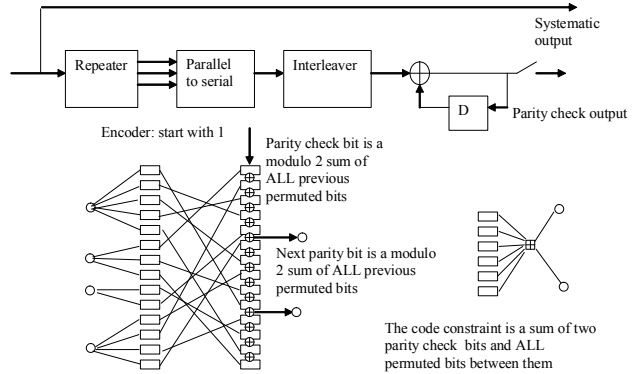

Fig.1: The basic RA encoder structure

## II. Decoding of IRA codes

The graph-decoding concept first proposed by [11], and researched by [12,13,14,15] includes the following. Decoding is performed iteratively, i.e., the purpose of the iteration is providing a priory information for the next iteration. Hence, the search for the nearest valid code word is performed by the sequential approximation method. Following [12], we can represent such code as a bipartite graph, which has two types of nodes: a variable node and a check node. The variable node represents a data bit or a soft decision obtained from a channel, and the check node represents a code constraint of a linear block code that must be always equal to 0. We briefly show the message-passing decoding algorithm in (Fig. 2) and refer the reader to [11] for more thorough discussion.

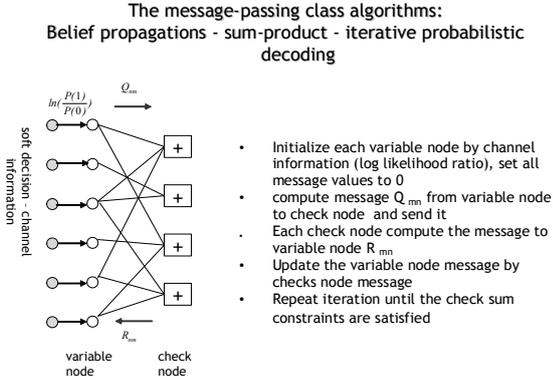

Fig. 2: Message passing algorithm

As it has been discussed the repeat accumulate code can be represented as the simplest convolutional code implemented in the trellis pattern having two states. The trellis diagram corresponding to this case is presented in Fig. 3.

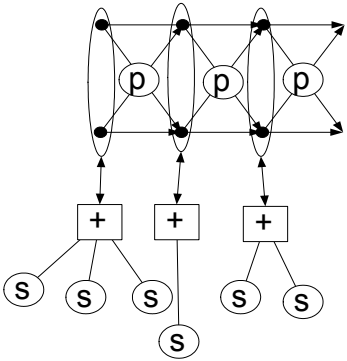

Fig. 3: Two state trellis of IRA codes.

Implementation of BCJR decoding to the RA codes features by the state probabilities of the MAP algorithm are involved in the message exchange process. The difference from the message-passing algorithm is that the check nodes of this case are updated sequentially, one by one. Thus, the updated outgoing message is processed immediately on the next parity check node as an incoming message without expecting for the next iteration. Forward and backward transitions are performed in the same manner as of the original MAP algorithm. This decoding algorithm is sequential, it is slower than the fully parallel message passing algorithm, but it allows reducing number of iterations or improving decoding efficiency. However, this algorithm is not conventional BCJR, which considers all possible states. The described algorithms can be stated as two versions of the belief propagation algorithm distinguishing by the use of LDPC-like scheduling and turbo-like scheduling [16].

III. GRAPHS AND INTERLEAVERS

Therefore, operations on the Tanner graph being a graphical model of code division to subcodes could be described in the form of repetition, interleaving and grouping. Let the number of edges connecting to the node be a degree of this node. The degree of the variable node is the number of data bit repetitions, and the degree of the check node is the number of parity checks in this node. The term "irregular repeat accumulate code" usually means a left irregular code. Let us assume that all considered codes are right regular or slightly irregular. It is shown in [10] that irregular repeat accumulate codes outperform regular repeat accumulate codes, and optimum node degree distribution is found for infinite blocks. The desired distribution is represented as a polynomial of the formal parameter $x$ such that $\lambda(x) = \sum \lambda_i x^i$, where the polynomial coefficient is obtained as an expansion, in the Taylor series, of the function derived from successful decoding condition [17,18]. Appropriate repetition degree distribution looks as follows (for example):

$$\lambda(x) = 0.2x^3 + 0.4x^7 + 0.1x^{11} + 0.15x^7 + 0.1x^{29} + 0.05x^{31} \quad (1)$$

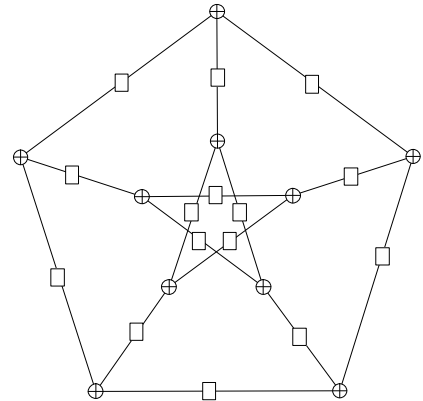

Fig. 4: A code on Peterson graph.

We will use this distribution in an AWGN channel for a finite, possibly very short block length. As an illustration, Fig. 4 presents the code constructed by Tanner on the Peterson graph with 6 in the diameter and the girth of 10. However, disadvantage of this approach is hardness of construction of a graph having the size of about 100 nodes and the specified properties.

IV. THE INTERLEAVER DESIGN BASED ON GRUENBAUM GRAPH

We proposed to use the known node degree distribution with a specially designed interleaver. The simplest interleaving rule (except rectangular) is known as a

relative prime interleaver and consists in the following permutation:

$$\Pi(i) = (p \times i + s) \bmod n \quad (2)$$

where the numbers $p, n$ are coprimes (the greatest common divisor is equal to 1) [19]. This method could be updated with dithered permutations [20]. The dithered permutation consists in that all symbols in a block with the length $n$ are divided to groups where the interleaving operation is executed by the specified (possibly, pseudorandom) interleaver having the length much less than $n$.

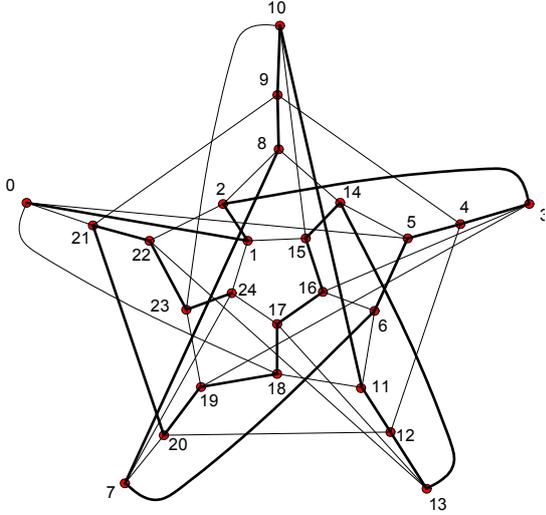

Fig. 5: Gruenbaum graph with Hamiltonian path.

We propose to obtain dithered sequences from the graphs with simultaneous possible largest girth $g$ and chromatic number $\gamma$. For $g = 5, \gamma = 4$, only one $\gamma$-regular graph is known. That is a Gruenbaum graph (Fig. 5). One can derive the dithered sequence from this graph by

1. Finding a path passing through each node once (Hamiltonian path) in the graph and enumerating all nodes in the order of passing. Black edge rout of Fig. 5
2. Passing across the remaining edges and storing numbers of the passed nodes. The dithered sequence is thus formed.

For the paths shown in Fig. 5 the resulting sequence is given in (3).

$$Gr_{25} = \begin{Bmatrix} 7,20,12,4,9,21,0,18,11,6, \\ 16,3,19,23,10,15,1,24,17,13,22,2,8,14,5 \end{Bmatrix}$$
(3)

This sequence has a triangle "s-random" property defined in [19]. It means that $|i - j| + |\Pi(i) - \Pi(j)| \geq S$ where $S = 5$, $i, j$ numbers before interleaving and $\Pi(i), \Pi(j)$ numbers after interleaving. It is important that the sequence does not include any periodical structure but is produced by deterministic operations. The final operation of the interleaver synthesis is the cyclic shift with overlapping in each group.

## V. SIMULATION

Fig. 6 shows the achieved gain with the designed interleaver of the IRA code over Viterbi decoding with 192 bits frame in the AWGN channel with R=1/4 in the turbo-like scheduling. The number of iterations is equal to 72. The interleaver length is $1344 = 192 \times 7$. The irregular repetition (left degree distribution) is defined by (1). The optimized numbers of $p = 173$ and $s = 1184$ have been found through exhaustive search. The search procedure contains finding cycles-4 in the Tanner graph with the involved variable nodes of minimal degree and finding stopping sets of minimal degree. The number of such defects should be minimized or avoided. Each convolutional codeword has 8-bit zero tail which finalizes the frame. Each IRA codeword has eight predetermined bits at the specified positions where the matrix defects are detected. The positions where "ones" are put have the following locations $\{3,9,11,18,19,26,27,74\}$.

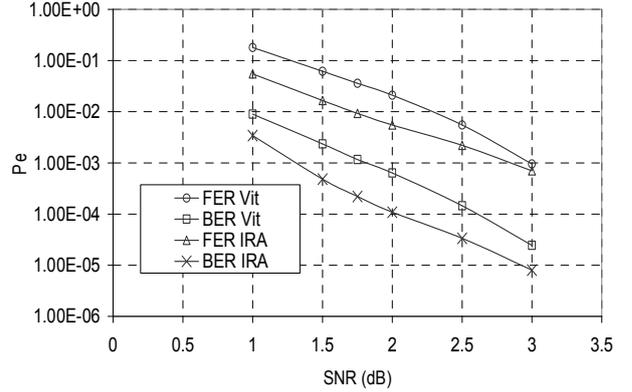

Fig. 6: Comparison of IRA and convolutional ($K = 9$) codes.

Actually 24-bit length sequence $Gr_{24}$ is used as a small interleaver.

$$Gr_{24} = \begin{Bmatrix} 7,20,12,4,9,21,0,18,11,6, \\ 16,3,19,23,10,15,1,17,13,22,2,8,14,5 \end{Bmatrix}$$
(4)

This array is obtained from $Gr_{25}$ sequence by excluding the number $24$. Note, that the small interleaver size should divide the whole interleaver size. The small sized interleavers has been proposed for turbo codes in [21,22] but such methods defined as permutation polynomial-based interleaving [22] extensively exploit the specific property of the convolutional turbo codes. The proposed method exploits a property of the IRA code and at the same time it follows the guideline of the early method: to develop a deterministic method of easy

(possibly on fly) interleaver generation which does not require additional memory to store interleaver coefficients. The algorithm implementation was shown in Fig. 7 in details.

```
void Gruenbaum_interleaver(int *dst){
    int length = 1344;    int len_I = 24;   int s=1184;
    int p=173;    int ptr[1344];   int ptr2[1344];
    I[] = {0,14,9,22,18,2,15,5,10,17,4,13,7,1,21,12,16,23,6,19,11,3,8,20};
    int i,j;
    for(i=0;i<length;i++){
        j = i%len_I;
        ptr[i] = (i-j) + I[j];
        ptr2[i]=0;
    }
    for(i=0;i<length;i++){
        dst[i] = (s+ptr[i]*p)%length;
        ptr2[(s+i*p)%length]++;
    }
    j=-1;
    for(i=0;i<length;i+=len_I){
        if(j>=0){
            int tmp = dst[j];
            dst[j]=dst[i];
            dst[i]=tmp;
        }
        j=i;
    }
}
```

Fig. 7 The interleaver design based Gruenbaum graph.

## VI. CONCLUSION

We have proposed new designs for the IRA code interleavers that are comparable with the Viterbi decoder by complexity and provide gain in the AWGN channel with the coding rate $R = 1/4$. The achieved gain is 0.7 - 0.8. at the level of 1 % FER. For decoding the message passing BCJR algorithm is used which is not the maximum likelihood algorithm in the case of finite block length codes but the achieved gain (especially in terms of BER) allows proposing this algorithm for channel coding of voice data.